# Autoignition of *n*-Butanol
# at Elevated Pressure and Low to Intermediate Temperature

Bryan W. Weber, Kamal Kumar, Yu Zhang, Chih-Jen Sung

Department of Mechanical Engineering

University of Connecticut

Storrs, CT 06269

Corresponding Author:

Chih-Jen Sung

Department of Mechanical Engineering

University of Connecticut

Room 484, United Technologies Engineering Building

Storrs, CT 06269, USA

Phone: (860) 486-3679

Fax:     (860) 486-5088

Email: cjsung@engr.uconn.edu

**Abstract**

Autoignition experiments for *n*-butanol have been performed using a heated rapid compression machine at compressed pressures of 15 and 30 bar, in the compressed temperature range of 675–925 K, and for equivalence ratios of 0.5, 1.0, and 2.0. Over the conditions studied, the ignition delay decreases monotonically as temperature increases, and the autoignition response exhibits single-stage characteristics. A non-linear fit to the experimental data is performed and the reactivity, in terms of the inverse of ignition delay, shows nearly second order dependence on the initial oxygen mole fraction and slightly greater than first order dependence on initial fuel mole fraction and compressed pressure. Experimentally measured ignition delays are also compared to simulations using several reaction mechanisms available in the literature. Agreement between simulated and experimental ignition delay is found to be unsatisfactory. Sensitivity analysis is performed on one recent mechanism and indicates that uncertainties in the rate coefficients of parent fuel decomposition reactions play a major role in causing the poor agreement. Path analysis of the fuel decomposition reactions supports this conclusion and also highlights the particular importance of certain pathways. Further experimental investigations of the fuel decomposition, including speciation measurements, are required.

1. Introduction

Recent concerns over energy security and the environment have created a renewed push to reduce our dependence on fossil fuels. Alternative fuels such as ethanol are replacing some of the petroleum-based fuels currently in use, especially in the transportation sector. For many reasons, however, ethanol is considered to be a poor replacement for current fuels. It is fully miscible in water, preventing the use of current fuel transportation infrastructure to distribute ethanol. It has a lower energy density than gasoline, which limits the amount that can be mixed into gasoline without modifying current engine technology. Finally, there are many concerns about the source of raw materials used to produce ethanol. Frequently, these materials are grown using environmentally harmful fertilizers and they often displace food crops, which may drive up the cost of food [1].

For these reasons, a new generation of alternative fuels is being developed. One of the most promising of these fuels is *n*-butanol, which ameliorates many of the concerns with the use of ethanol [2]. *n*-Butanol can be mixed with gasoline in much higher proportions due to its higher energy content (28.4 MJ/L for *n*-butanol vs. 21.2 MJ/L for ethanol [3]), which more closely matches that of gasoline (31.9 MJ/L [4]). It is more hydrophobic than ethanol, which allows the use of currently existing fuel distribution infrastructure. In addition, *n*-butanol can be produced by fermentation from many sources, including waste products, such as corn stover, and crops that do not displace food crops. Significant work has been done recently to genetically add the ability to produce *n*-butanol to common industrial bacteria, such as *E. Coli* [5-10]. Other researchers are working to increase the production efficiency of *n*-butanol by organisms which innately contain the ability to produce *n*-butanol [11,12]. Several excellent review articles covering the work on bioproduction of *n*-butanol are also available (c.f. Refs. [8,10,13]).

Until recently, very little work on the combustion of *n*-butanol had been performed. In the 1950's, Barnard performed pyrolysis studies of *n*- and *tert*-butanol [14,15], while Smith et al. [16] performed sampling of the diffusion flames of the four isomers of butanol. In the 1960's, Cullis and Warwicker [17] performed work on the "slow-combustion" of the butanol isomers and reported cool flames of *n*-butanol between 305 and 340 °C. In the 1980's, Hamins [18] used a diffusion flame setup to report extinction limits of *n*-butanol and *n*-butanol/heptanes/toluene mixtures. In the last 15 years, the pace of research has started to increase. Recent works include gasoline and diesel engine studies [3,4,19-27], jet-stirred reactor (JSR) studies [28-31], low-pressure flame sampling studies using EI- and PI-MBMS to identify combustion intermediates [32,33], measurements of laminar flame speeds [31,34-36], sampling measurements from diffusion flames [31,37,38], ignition delay measurements in shock tubes [39-41], and pyrolysis studies [42].

Among the shock tube ignition studies, Moss et al. [39] have done measurements for all four isomers of butanol at 1 and 4 bar and 1200–1800 K, over equivalence ratios of $\phi$=0.5, 1, 2 and fuel mole percents of 0.25%, 0.5%, and 1%. Black et al. [40] investigated autoignition for *n*-butanol from 1100–1800 K and 1, 2.6, and 8 atm over $\phi$=0.5, 1, 2 and fuel mole percents of 0.6%, 0.75% and 3.5%. Further, Heufer et al. [41] reported high pressure ignition delay results of stoichiometric *n*-butanol/air mixtures under the conditions behind the reflected shock of approximately 10–42 bar and 770–1250 K. Specifically, the results of Heufer et al. [41] showed an interesting non-Arrhenius behavior at temperatures lower than about 1000 K for the pressure range studied. They found that the rate of increase of ignition delay with decreasing temperature appears to change around 1000 K. They also presented a sensitivity analysis of the reaction mechanism developed by Black et al. [40] and found that a fuel + $HO_2$ reaction is among the most sensitive, especially in the low temperature regime.

Although the works mentioned previously represent a wide body of knowledge of *n*-butanol combustion, the amount of work that has been conducted on the autoignition of *n*-butanol at elevated pressures is still rather limited. The goal of this study is to provide additional autoignition data of *n*-butanol at elevated pressures and low-to-intermediate temperatures. Using a rapid compression machine (RCM), ignition delays for high pressure conditions of 15–30 bar and low-to-intermediate temperature conditions of 675–925 K are obtained. Experimental results are also modeled using several reaction mechanisms available in the literature. Furthermore, based on one recent mechanism, several methods of analysis are used to identify the controlling chemistry of *n*-butanol autoignition under the conditions relevant to this study. According to analysis of model results and experimental data, future studies needed for mechanism refinement are discussed.

## 2. Experimental and Computational Specifications

### 2.1 Rapid Compression Machine

Autoignition delay measurements are performed in a rapid compression machine (RCM). The RCM compresses a fixed mass of premixed mixture of fuel and oxidizer to a given temperature and pressure over about 25–35 milliseconds. The compression ratio, initial temperature, and initial pressure are adjusted to vary the compressed temperature of the gas mixture at constant compressed pressure. This means that, in general, the uncertainty in the compressed temperature is dependent on the initial conditions of the experiment, which will be discussed in due course. The piston used to compress the gases is machined with crevices designed to suppress the roll-up vortex effect and ensure a homogeneous reaction zone. The reaction chamber is fitted with pressure and temperature sensing devices to measure the initial conditions in the reaction chamber. Additionally, the reaction chamber is fitted with a Kistler 6125B dynamic pressure transducer to measure the pressure in the reaction chamber during compression and any post-compression events including ignition. Further details of the RCM are available in Ref. [43].

**2.2 Mixture Preparation**

*n*-Butanol (anhydrous, 99.9%), $O_2$ (99.8%), and $N_2$ (99.998%) are used as the reactants. To determine the mixture composition, the mass of fuel, equivalence ratio ($\phi$), and oxidizer ratio ($X_{O2} : X_{inert}$, where $X$ indicates mole fraction) are specified. Since *n*-butanol is liquid at room temperature and has a relatively low vapor pressure (~6 Torr at 25 °C), it is measured gravimetrically in a syringe to within 0.01 g of the specified value. Proportions of $O_2$ and $N_2$ in the mixture are determined manometrically and added at room temperature. The saturated vapor pressure dependence of *n*-butanol on temperature is taken from the *Chemical Properties Handbook* by Yaws [44]. The preheat temperature is set above the saturation temperature of *n*-butanol to ensure complete vaporization of the fuel. A magnetic stirrer mixes the reactants. The temperature inside the mixing tank is allowed approximately 1.5 hours to reach steady state.

**2.3 Mixture Composition Check**

Tests with GCMS are conducted to check that the expected mixture is present in the mixing tank for the entire duration of experiments. A mixture is prepared exactly as previously described, except a known concentration of *iso*-octane is added. This functions as an internal standard from which the concentration of *n*-butanol can be calculated. Figure 1(a) shows the overall chromatogram of the separation of a sample withdrawn from the mixing tank, while Fig. 1(b) is an enlargement of Fig. 1(a) from retention time of 12.5 to 17 minutes.

The first objective of the mixture composition check is to ensure there is no decomposition of *n*-butanol during preheating in the mixing tank. As a reference, Fig. 1(c) shows the chromatogram of a liquid calibration sample consisting of *n*-butanol and *iso*-octane (Sigma-Aldrich, 99.9%) diluted in acetone (Sigma-Aldrich, 99.5%). The two small peaks in Fig. 1(c) are impurities in the *iso*-octane and *n*-butanol; ethyl acetate is from *n*-butanol and 2,3-dimethylpentane is from *iso*-octane. During the mixture composition check, the mixing tank is heated to 87 °C and contains a $\phi=0.5$ mixture, with 1.1% by mole *iso*-octane replacing an equivalent amount of nitrogen. It is seen from Fig. 1(b) that no additional peaks are present as compared to Fig. 1(c), indicating there is no decomposition of *n*-butanol.

The second objective of the mixture composition check is to verify that the concentration of *n*-butanol matches the expected value. The expected value is calculated as the mole fraction of *n*-butanol in a $\phi=0.5$ mixture in air. The response factor of *n*-butanol to *iso*-octane in the liquid sample is calculated based on species peak area ratio in Fig. 1(c) and the known concentrations [45]. Then, samples from the mixing tank are withdrawn and analyzed using the GCMS. The peak area of each component is calculated, and using the known concentration of *iso*-octane and the response factor, the concentration of *n*-butanol is determined. A total of five samples from one mixing tank are analyzed. The concentration of *n*-butanol is found to be within 4% of the expected value for this representative case. Based on these results, it is confirmed that the present mixture preparation procedure is adequate for obtaining a homogeneous mixture.

## 2.4 Experimental Conditions

Experiments are carried out under a wide variety of conditions. The compressed pressure ($P_C$) conditions chosen have not been covered extensively by previous work. The fuel loading conditions are also chosen to cover ranges not studied in previous work. Most of the shock tube studies have used relatively dilute mixtures – only 1–3 percent by mole of fuel. This study includes conditions at twice the maximum fuel concentration of previous work. Furthermore, at equivalence ratios of 0.5 and 2.0, the concentrations of fuel and oxygen are separately adjusted to reveal the effect of each on the ignition delay. Table 1 shows the experimental conditions studied in this work. Further, the compressed temperatures investigated are in the range of $T_C$=675–925 K, over which the literature ignition delay data are meager.

## 2.5 Experimental Reproducibility

Each compressed pressure and temperature condition is repeated at least 6 times to ensure reproducibility. The mean and standard deviation of the ignition delay for all runs at each condition are calculated; as an indication of reproducibility, the standard deviation is less than 10% of the mean in every case. Representative experimental pressure traces for simulations and plotting are then chosen as the closest to the mean. Figure 2 shows a set of runs from a representative condition. In addition, each new mixture preparation is checked against previously tested conditions to ensure consistency.

## 2.6 Simulations and Determination of Compressed Temperature

Two different types of simulations are performed using CHEMKIN-PRO [46] and SENKIN [47] coupled with CHEMKIN-III [48]. There are no substantial differences in the results obtained from these programs. The first type of simulation is a constant volume, adiabatic simulation, whose initial conditions are set to the pressure and temperature in the reaction chamber at top dead center (TDC). The second type includes both the compression stroke and post-compression event by setting the simulated reactor volume as function of time. Heat loss during and after compression are modeled empirically to fit the experimental pressure trace of the corresponding non-reactive counterpart, as described in Refs. [43,49-53]. The literature reaction mechanisms used in simulations are taken from Grana et al. [38], Moss et al. [39], Black et al. [40], and Harper et al. [42].

Temperature at top dead center is taken as the reference temperature for reporting ignition delay data and is obtained from the RCM simulations at the end of compression; this is reported as the compressed temperature ($T_C$). This approach requires the assumption of an "adiabatic core" of gases in the reaction chamber, which is facilitated on the present RCM by the creviced piston described previously. When the same temperature is reached at the end of compression whether reactions are included in the simulation or not, this indicates there is no major chemical heat release during the compression stroke. This approach has been validated in Refs. [43,49-54].

## 2.7 Definition of Ignition Delay

The end of compression, when the piston reached TDC, is identified by the maximum of the pressure trace prior to the ignition point. The local maximum of the derivative of the pressure with respect to time, in the time after TDC, is defined as the point of ignition. The ignition delay is then the time difference between the point of ignition and the end of compression. Figure 3 illustrates the definition of ignition delay (τ) used in this study, where $P(t)$ is the pressure trace and $P'(t)$ is the time derivative of the pressure trace.

## 3. Results and Discussion

### 3.1 Experimental Results

Figure 4 shows one of the key features of the RCM, namely, the ability to vary compressed temperature at constant compressed pressure. The conditions in this figure are representative of conditions in all the experiments. As seen in Fig. 4, ignition delay decreases monotonically as compressed temperature increases, indicating that these experiments are not in the negative temperature coefficient (NTC) region. Indeed, no NTC region is observed for the conditions investigated in this study. It is noted that *n*-butane [55] and propane [56] exhibit a negative temperature dependence in a similar temperature and pressure range, but the autoignition of *n*-propanol has not, to the authors knowledge, been studied in the present temperature and pressure range to determine if a comparison can be drawn [57,58].

It is also clear from Fig. 4 that two-stage ignition did not occur. This is the case for all the conditions studied in this work and agrees with the work of Zhang and Boehman [25]. However, $C_3$ and greater alkanes exhibit two-stage ignition in the temperature and pressure range studied [55,59]. Again, *n*-propanol has not been studied in this temperature and pressure range to determine if a comparison is suitable. Finally, although Cullis and Warwicker [17] reported cool flames of *n*-butanol at low temperatures (575–625 K), their work was also at very low pressure (100–240 Torr) and, therefore, is not directly comparable to the current, high-pressure, results.

Figure 5 shows a comparison of the ignition delays for three equivalence ratios, $\phi$=0.5, 1.0, and 2.0, made in air. The vertical error bars are two standard deviations of the ignition delay. The standard deviation is computed based on all the runs at a particular compressed temperature and pressure condition. The lines are least squares fits to the data.

In general, the uncertainty in the compressed temperature is dependent on the measurement of the initial pressure and temperature, as well as the measurement of the compressed pressure. According to the manufacturer's specifications, the error of the initial temperature measurement is ±2 K, while the error of the initial pressure measurement is ±0.625 psi and the error of the compressed pressure measurement is ±1%. A detailed uncertainty analysis has been conducted to determine the contributions of these factors to the uncertainty of the calculation of the compressed temperature. The largest contributor to the total uncertainty in compressed temperature is the error in the initial pressure measurement, whereas the contributions from the measurements of the initial temperature and the compressed pressure are less significant. In addition, the uncertainty is dependent on the actual value of the initial pressure, but not the actual values of the initial temperature and compressed pressure. Typically, higher initial pressures result in lower uncertainties in the compressed temperature. Taken all together, a conservative estimate of the total uncertainty in the compressed temperature is about 0.7% to 1.7% of the compressed temperature. Reducing this uncertainty can be accomplished most simply by improving the accuracy of the pressure transducer used to measure the initial pressure.

The effect of equivalence ratio on the ignition delay is quite clear – increasing the equivalence ratio increases the reactivity in the temperature range investigated. That is, similar ignition delays are found at lower temperatures as the equivalence ratio increases. It is noted from Fig. 5 that for *n*-butanol/air mixtures the reactivity dependence on the equivalence ratio can change in different temperature ranges. This is evident from the lines through the experimental data, which appear to intersect around 850–900 K. By extending the trend to higher temperatures, the lower equivalence ratios could become more reactive than the higher equivalence ratios. The high-temperature shock tube studies of Moss et al. [39] and Black et al.

[40] support this conclusion, although their data are at lower pressures and lower fuel loading conditions. In the work of [39,40], the lower equivalence ratios have higher reactivity than the higher equivalence ratios.

The dependence of ignition delay on equivalence ratio is corroborated by the results shown in Fig. 6(a), which shows ignition delay as the fuel mole fraction varies with constant oxygen mole fraction. Again, the lowest fuel mole fraction is the least reactive, and increasing the fuel mole fraction increases reactivity. A similar effect of oxygen concentration is shown in Fig. 6(b). Increasing the oxygen concentration increases reactivity at constant fuel mole fraction. A notable observation from Figs. 5 and 6 is the rather large changes in the slope of the ignition delay plot with equivalence ratio. This may be due to a change in the controlling autoignition chemistry within the temperature range of this study.

Figure 7 shows comparison of data at $P_C$=15 and 30 bar and $\phi$=1.0 in air. Again, the solid lines through the experiments are least squares fits to the data. The 30 bar cases are more reactive than the 15 bar cases, as is expected.

Figure 8 shows comparison of the shock tube data of Heufer et al. [41] and the data from the current RCM experiments. In general, there is quite good agreement, and the data of Heufer et al. [41] appears to be consistent with the current data. That is, the highest pressures achieve the highest reactivity, and they are approximately in line with the results from this study.

A non-linear regression is further performed to determine a correlation of ignition delay with reactant mole fractions and compressed pressure. This correlation is given as follows:

$$\tau = 10^{-(8.5\pm0.8)} X_{O_2}^{-(1.7\pm0.2)} X_{n-Butanol}^{-(1.4\pm0.2)} P_C^{-(1.5\pm0.3)} \exp[(9703.3 \pm 1035.5)/T_C] \text{ (seconds)},$$

where $X_i$ is the mole fraction of species $i$ and $P_C$ and $T_C$ are in units of bar and K, respectively. The uncertainties in the correlation parameters are the uncertainties in the regression resulting from scatter in the data, and they result in a typical uncertainty in the prediction of $\tau$ of ±5 ms. The result of this regression analysis is shown in Fig. 9. One data set, ϕ=0.5 in air, appears to deviate from the trendline in Fig. 9. The cause of this noticeable deviation is probably related to the change in slope discussed in connection with Fig. 5. Further, this correlation is calculated over the range of temperature and pressure investigated herein and does not include data from the literature. Thus, extending the correlation to temperature or pressure ranges far outside the current work is unlikely to yield acceptable results. All of the ignition delay data from this study can be found in the supplementary material.

**3.2 Reaction Mechanism Analysis**

Figure 10 shows a comparison of simulations with the experimental pressure trace at a representative condition. For all conditions in this study, none of the mechanisms tested are able to reproduce the ignition delay within the experimental duration, despite matching the pressure trace during the compression stroke and post-compression, pre-ignition event very well. Therefore, constant volume, adiabatic simulations are shown in the other figures in this work.

As shown by Fig. 11, even constant volume adiabatic simulations of ignition delay are much longer than the experimentally measured ignition delays, and at every condition studied here simulation of the ignition delay significantly over-predicts the experimental ignition delay. Note the y-axis of Fig. 11 is labeled in seconds, not milliseconds. It has to be pointed out, however, that none of the mechanisms have been validated for ignition delays in this temperature and pressure regime. It appears from Fig. 11 that the simulations and experiments would merge as the temperature increases. Nevertheless, the degree of difference between the simulations and the experiments is quite surprising. A more detailed analysis of the mechanism of Black et al. [40] is carried out to determine the major causes of this discrepancy.

### 3.2.1 Brute Force Sensitivity Analysis

First, brute force sensitivity analysis is conducted using constant volume, adiabatic simulations. Percent sensitivity is defined by the following:

$$\% \text{ sensitivity} = \frac{\tau(2k_i) - \tau(k_i)}{\tau(k_i)} \times 100\%,$$

where $k_i$ is the rate of reaction $i$, $\tau(2k_i)$ is the ignition delay when the rate of reaction $i$ has been doubled, and $\tau(k_i)$ is the nominal value of the ignition delay. Thus, a positive value for sensitivity means that the ignition delay becomes longer when the rate of reaction $i$ is doubled. Figure 12 illustrates the results of this analysis, at 15 atm and $\phi$=1.0 in air, over a range of initial temperatures. Three initial temperatures are chosen: 1800 K and 1100 K correspond to the upper and lower temperatures for which this mechanism has been validated, while 700 K is representative of the conditions in this study.

Figure 12 shows the ten most sensitive reactions from the 1100 K case, along with the sensitivity of the same reactions at 700 K and 1800 K. Key radicals such as O, H, OH, $HO_2$, and $H_2O_2$ are important in ignition processes of hydrocarbon fuels [60]. Therefore, it is not surprising to see many reactions among those radicals in the results. The other reactions, however, involve the initial decomposition of the fuel. At low temperature, the system is clearly most sensitive to reaction (R1):

$$n\text{-}C_4H_9OH + HO_2 \leftrightarrow C_4H_8OH\text{-}1 + H_2O_2, \tag{R1}$$

which is H-abstraction from the $\alpha$-carbon of $n$-butanol and is the same reaction found to be quite sensitive by Heufer et al. [41].

At low temperatures, H-abstraction from the fuel plays a major role in the combustion process [59,60]. Zhang and Boehman [25] also showed that H-abstraction is important in low temperature *n*-butanol ignition. Due to the effect of the alcohol group, the $\alpha$-carbon has the most weakly bonded hydrogens in *n*-butanol [40]. Therefore, it is reasonable to expect the system to be very sensitive to reaction (R1) at low temperature. By increasing the temperature, the sensitivity of the system to this reaction is reduced, such that, at 1800 K, the sensitivity is less than 1%. This indicates that increasing the rate of this reaction may improve simulation results at low temperatures but will not strongly affect correlation with existing high temperature data for ignition delay.

A similar analysis is also conducted at 800 K while varying the pressure from 1 to 30 atm. The results of this analysis show that the sensitivity of most reactions is similar over the entire range, suggesting that changing the rate of reaction (R1) will not negatively impact correlation with previous results.

Ignition delays over a large temperature range are computed for the several multiples of the rate coefficient of reaction (R1). It is found that even for low multiplication factors, the ignition delay becomes shorter over the entire temperature range investigated. However, agreement with low temperature experiments is improved significantly as the reaction rate of (R1) is increased, while agreement at high temperature is not significantly decreased.

Finally, Fig. 13 plots a comparison of the rate coefficient for reaction (R1) over the temperature range of 500–2000 K at one atmosphere. It can be seen that the rate coefficient in the mechanism of Harper et al. [42] has the lowest value over nearly the entire temperature range presented here. In spite of this, it can be seen in Fig. 11 that the mechanism of Harper et al. [42] best predicts the ignition delay. This suggests there are several important pathways of *n*-butanol oxidation that need to be accounted for, and trivial changes to the mechanisms will not necessarily bring them in line with the experiments.

### 3.2.2 Reaction Path Analysis

The second method of mechanism analysis is reaction path analysis. A reaction path diagram is one way of visualizing this analysis. The reaction path diagram shows the percent of each reactant destroyed to form the product indicated by the arrow. The percent destruction represents the cumulative destruction of each reactant up to the point in time where the mole fraction of the fuel is reduced 20% compared to the initial mole fraction. The time of 20% fuel consumption is chosen for two reasons. First, it has been used previously in the literature. Second, it is before the point of thermal runaway when the controlling chemistry changes. Note that the following analysis applies to the original mechanism of Black et al. [40], and again constant volume, adiabatic simulations are conducted.

Reaction path analysis is performed at two initial temperatures (800 K and 1600 K), with the same initial pressure (15 atm) and equivalence ratio ($\phi=1.0$ in air). Figure 14 shows the first level of *n*-butanol decomposition. The percent destruction values for the cases of 800 K and 1600 K are indicated in plain text and bold, italicized text, respectively. It can be seen that unimolecular decomposition destroys none of the fuel in the low temperature case, but a significant fraction at high temperature. This is expected, as unimolecular decomposition reactions are characterized by rather high activation energy. Figure 14 also illustrates that H-abstraction from the $\alpha$-carbon destroys the largest percent of fuel in the low temperature case. This is also an expected result, as discussed previously.

Figure 15 shows the reaction path diagram for two pathways of fuel decomposition in the 800 K case. The left pathway ends with the formation of but-1,3-en-1-ol. This pathway is important because it consumes nearly 30% of the fuel, by far the largest fraction. But-1,3-en-1-ol is a diene, and is accumulating in the system up to the point of 20% fuel consumption. That is, the total production of this product exceeds its destruction. In addition, in the mechanism of Black et al. [40], this species can only be formed from $C_4H_6OH$ by unimolecular fission of a C-H bond. However, according to Walker and Morley [61], the primary route of formation of dienes from 1-butene and 2-butene is by reaction with molecular oxygen, to form the diene and $HO_2$. This pathway is supported by Vanhove et al. [62] who discussed general allyl decomposition pathways. They also noted the additional pathway of molecular oxygen addition followed by several isomerization steps, leading to the diene and $HO_2$.

Another reaction in this pathway which may be delaying the onset of ignition is the reaction of $C_4H_6OH$ and $HO_2$ to form but-1-en-1-ol and $O_2$, which destroys over 95% of $C_4H_6OH$. This step seems unlikely to be prominent, as it involves the release of an oxygen molecule. Indeed, at high temperature, this reaction is only a minor contributor, and the direction has been reversed; that is, the reaction is producing the resonant structure from the enol instead of the other way around. Part of this may have to do with the fact that $HO_2$ reactions are much less important at high temperatures; however, it is useful to emphasize that the reaction direction has been reversed.

The right pathway in Fig. 15 contains another important intermediate in *n*-butanol combustion, butanal. Butanal has been detected by several studies, including flames [32], jet-stirred reactors [31], and engine studies [25]. It can be formed by several pathways, detailed in the work of Zhang and Boehman [25] and Grana et al. [38]. It can be seen from Fig. 15 that, in the mechanism of Black et al. [40], the formation of butanal occurs primarily by attack of molecular oxygen on the alcohol group of $\alpha$-hydroxybutyl radicals, producing butanal and $HO_2$. The pathway of tautomerization of but-1-en-1-ol to butanal is not present in this mechanism.

Black et al. [40] reported that the simulated concentrations of but-1-en-1-ol were much higher than butanal. However, experimental results from the jet-stirred reactor [29] showed that the measured concentration of butanal was higher than the simulated profiles of both but-1-en-1-ol and butanal, and was in fact approximately equal to the sum of the simulated butanal and but-1-en-1-ol profiles. Black et al. [40] postulated that rapid tautomerization of but-1-en-1-ol to butanal occurred between the reaction chamber and the species analysis device, thus leading to the anomalous readings.

In contrast, Grana et al. [38] proposed that, although enols will be formed at high temperatures, tautomerization is so rapid that only the corresponding ketones will be present. This approach does not seem to agree with the results of the study by Yang et al. [32] who reported ionization signals at the ionization threshold of all of the butenols, propenols, and ethenol. Other works in the literature have shown that, although enols are a minor species in most flames, they do exist and need to be accounted for [63].

Furthermore, Zhang and Boehman [25] reported small concentrations of but-2-en-1-ol and but-3-en-1-ol and 300 times higher concentrations of butanal in their engine ignition study. They thus inferred that butanal was primarily formed directly from $\alpha$-hydroxybutyl, rather than primarily by tautomerization from but-1-en-1-ol. Harper et al. [42] supported this conclusion in their paper. When validating their mechanism against the JSR data of Sarathy et al. [31], Harper et al. [42] found that the dominant route to butanal formation was the reaction of $\alpha$-hydroxybutyl with molecular oxygen to form butanal and hydroperoxy radical, while minor pathways were tautomerization of but-1-en-1-ol, $\beta$-scission of $\alpha$-hydroxybutyl, and the reaction of $n$-butoxy radical with molecular oxygen.

In addition, Zhang and Boehman [25] showed that a significant amount of propanal formed prior to ignition. They suggested that the pathway responsible for the formation of propanal starts with molecular oxygen addition to the $\beta$-hydroxyalkyl radical of the fuel. Yang et al. [32] also showed propanal as a product of 1-butanol combustion, although they were not able to provide quantitative data in that study. This channel is entirely absent from the mechanism of Black et al. [40]. Propanal subsequently decomposes into either formaldehyde + $C_2H_5$ or ethenal + $CH_3$, both of which are important radical channels. Thus, adding this channel may improve simulation results.

Figure 16 shows the percent consumption of $\alpha$-hydroxybutyl by three major reactions as a function of multiplication factor of reaction (R1). The initial conditions of the simulation are 800 K, 15 atm, and $\phi$=1.0 in air. All other consumption reactions for $\alpha$-hydroxybutyl are less than one percent. It is interesting that as soon as the rate of reaction (R1) is increased, the percent consumed into the butanal pathway increases markedly and the pathway into $C_2H_3OH$ + $C_2H_5$ decreases markedly, while the percent directed into but-1-en-1-ol remains approximately the same. It is also of interest that, even as the rate of reaction (R1) is increased further, the percent of $\alpha$-hydroxybutyl consumed by each reaction remains approximately constant. This result shows the importance of the butanal pathway in the oxidation of $n$-butanol.

All together, these analyses suggest that it will be necessary to modify more than just one reaction to bring simulations in line with experimental results. It is apparent that the reactions of the $\alpha$-hydroxybutyl radical, both formation and destruction, are among the most important in the system, and a thorough investigation of these pathways is warranted. Quantum calculations can provide some insight into the reaction rates relevant to the decomposition of *n*-butanol. In addition, speciation measurements during the ignition process can help determine which pathways are most prominent in the fuel decomposition.

## 4. Conclusions

In this rapid compression machine study, autoignition delays of *n*-butanol are measured at low-to-intermediate temperatures and at elevated pressures. In particular, compressed temperature conditions of $T_C$=675–925 K are studied at compressed pressures of $P_C$=15 and 30 bar. Results demonstrate that higher pressure experiments have shorter ignition delays than lower pressure cases. In addition, independent variation of the concentration of fuel and oxygen at three equivalence ratios ($\phi$=0.5, 1.0, and 2.0) has revealed the effect of each on the ignition delay. Increasing the fuel and oxygen concentrations both decrease the ignition delay, while decreasing either increases the ignition delay. Of particular note is the lack of NTC and two-stage ignition delay in the temperature and pressure range studied. This result is of interest as all alkanes with three or more carbon atoms show NTC in the temperature range studied here.

Simulated ignition delays computed using four reaction mechanisms available in the literature are much longer than experimental ignition delays. In some cases, the discrepancy is several orders of magnitude. Although the reaction mechanisms have not been validated in the temperature and pressure range studied here, the degree of difference is still rather surprising.

To determine the source of the errors in the reaction mechanisms, the mechanism of Black et al. [40] is analyzed by two methods. The first, a brute force sensitivity analysis, shows that one reaction is much more important than the rest at low temperatures. This reaction is hydrogen abstraction from *n*-butanol by $HO_2$ to form the $\alpha$-hydroxybutyl radical. However, modifying the rate of just this reaction is insufficient to bring the simulations in line with the experiments. Second, a reaction path analysis is performed to reveal the pathways by which the fuel decomposes. This analysis reveals several pathways which may need to have their rates adjusted and at least two pathways which are missing entirely. These missing pathways are propanal formation and tautomerization of but-1-en-1-ol to butanal. Further experimental and theoretical studies on autoignition of *n*-butanol are warranted.

**Acknowledgements**

This material is based upon work supported as part of the Combustion Energy Frontier Research Center, an Energy Frontier Research Center funded by the U.S. Department of Energy, Office of Science, Office of Basic Energy Sciences under Award Number DE-SC0001198. B.W.W. was also supported by the Graduate Assistantship in Areas of National Need Pre-Doctoral Fellowship.

**Figure Captions**

Figure 1    (a) Overall chromatogram of *n*-butanol and *iso*-octane separation for $\phi$=0.5 mixture in air, at mixing tank temperature of 87 °C. (b) Enlargement of (a) from 12.5 to 17 minutes. (c) Chromatogram of the liquid sample used for calibration. Comparison of (b) and (c) shows no decomposition of *n*-butanol in the mixing tank at this preheat temperature.

Figure 2    Representative set of runs showing the reproducibility of experiments.

Figure 3    Definition of ignition delay used in this study. $P(t)$ is the pressure as a function of time and $P'(t)$ is the time derivative of the pressure, as a function of time.

Figure 4    Pressure traces at varying compressed temperatures for the experiments of $P_C$=15 bar and $\phi$=0.5 in air.

Figure 5    Ignition delay comparison of three equivalence ratios in air for compressed pressure of $P_C$=15 bar. Lines are least squares fits to the data.

Figure 6    (a) Comparison of three fuel concentrations at three equivalence ratios and $P_C$=15 bar. In these, the oxygen mole fraction is held constant. Lines are linear least squares fits to the data. (b) Comparison of three oxygen concentrations at three equivalence ratios and $P_C$=15 bar. In these, the fuel mole fraction is held constant. Lines are linear least squares fits to the data.

Figure 7    Comparison of experiments at compressed pressures of $P_C$=15 and 30 bar for $\phi$=1.0 in air.

Figure 8    Current ignition delay data compared to the data from Heufer et al. [41]. Filled symbols are RCM data from the current work; open symbols are shock tube data from Heufer et al. [41].

Figure 9    Correlation for measured ignition delays in RCM. "Air" indicates the mixture was made in air; "Fuel" indicates the initial fuel mole fraction was varied at constant initial oxygen mole fraction; "Oxygen" indicates the initial oxygen mole fraction was varied at constant initial fuel mole fraction.

Figure 10    Comparison of RCM simulations with experimental pressure trace.

Figure 11  Comparison of computed ignition delays using shock tube simulations and RCM experimental results. The line through the experiments is a least squares fit to the data. Note the y-axis is labeled in seconds, not milliseconds.

Figure 12  Results of brute force sensitivity analysis at three different temperatures for initial pressure of 15 atm and ϕ=1.0 in air.

Figure 13  Comparison of the literature rate coefficients for the reaction of *n*-butanol and hydroperoxy radical to form $\alpha$-hydroxybutyl and hydrogen peroxide.

Figure 14  Initial fuel decomposition (20% reduction in fuel mole fraction compared to the initial mole fraction) in the mechanism of Black et al. [40] at initial conditions of 15 atm and ϕ=1.0 in air. Plain text is for 800 K initial temperature, while bold, italicized text is for 1600 K initial temperature.

Figure 15  Pathway of but-1,3-en-1-ol formation, through but-1-en-1-ol. Also, the pathway of butanal formation is shown. The "**X**" through the arrow from but-1-en-1-ol indicates that pathway is not present in the mechanism. Initial conditions are 800 K, 15 atm, and ϕ=1.0 in air. Percent consumption is computed up to 20% reduction in fuel mole fraction compared to the initial mole fraction.

Figure 16  Percent consumption of α-hydroxybutyl by three key reactions at various multiplication factors of reaction (R1). Initial conditions are 800 K, 15 atm, and ϕ=1.0 in air. Percent consumption is computed up to 20% reduction in fuel mole fraction compared to the initial mole fraction.

**Table Captions**

Table 1  Experimental conditions studied in this work, with compressed temperatures in the range of $T_C$=675–925 K.

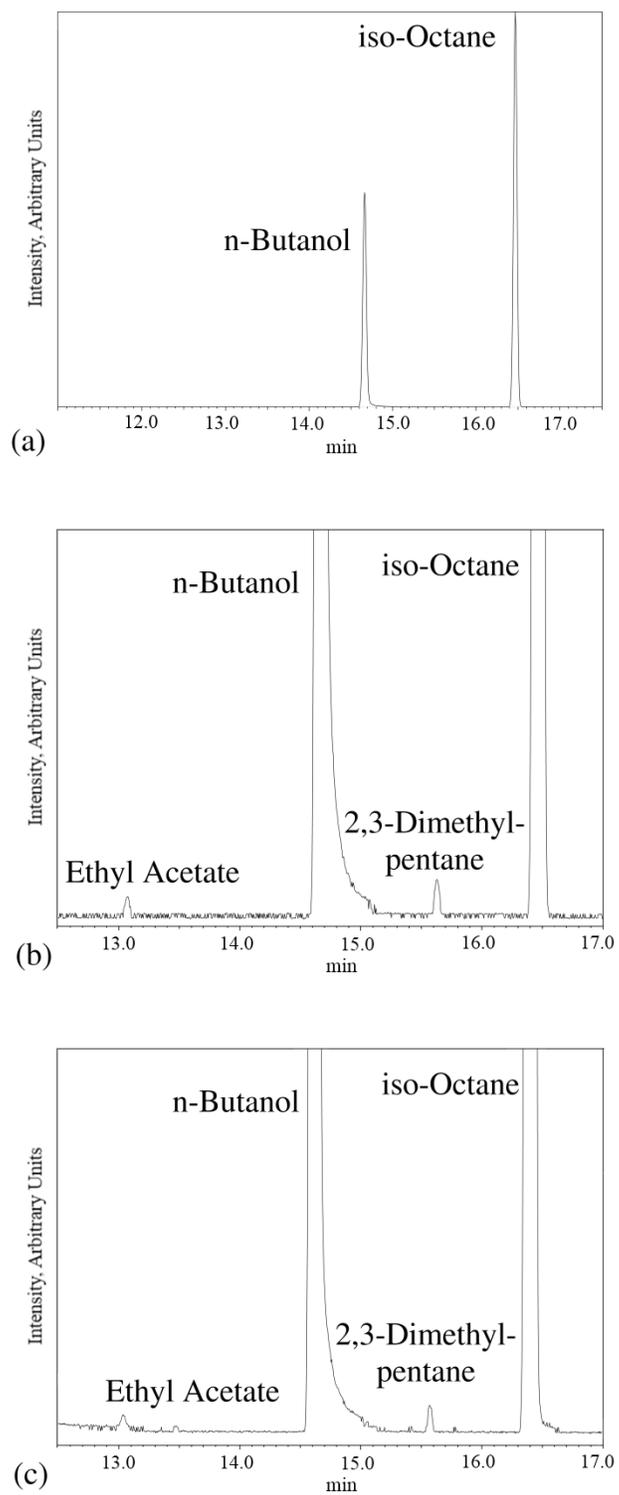

Figure 1

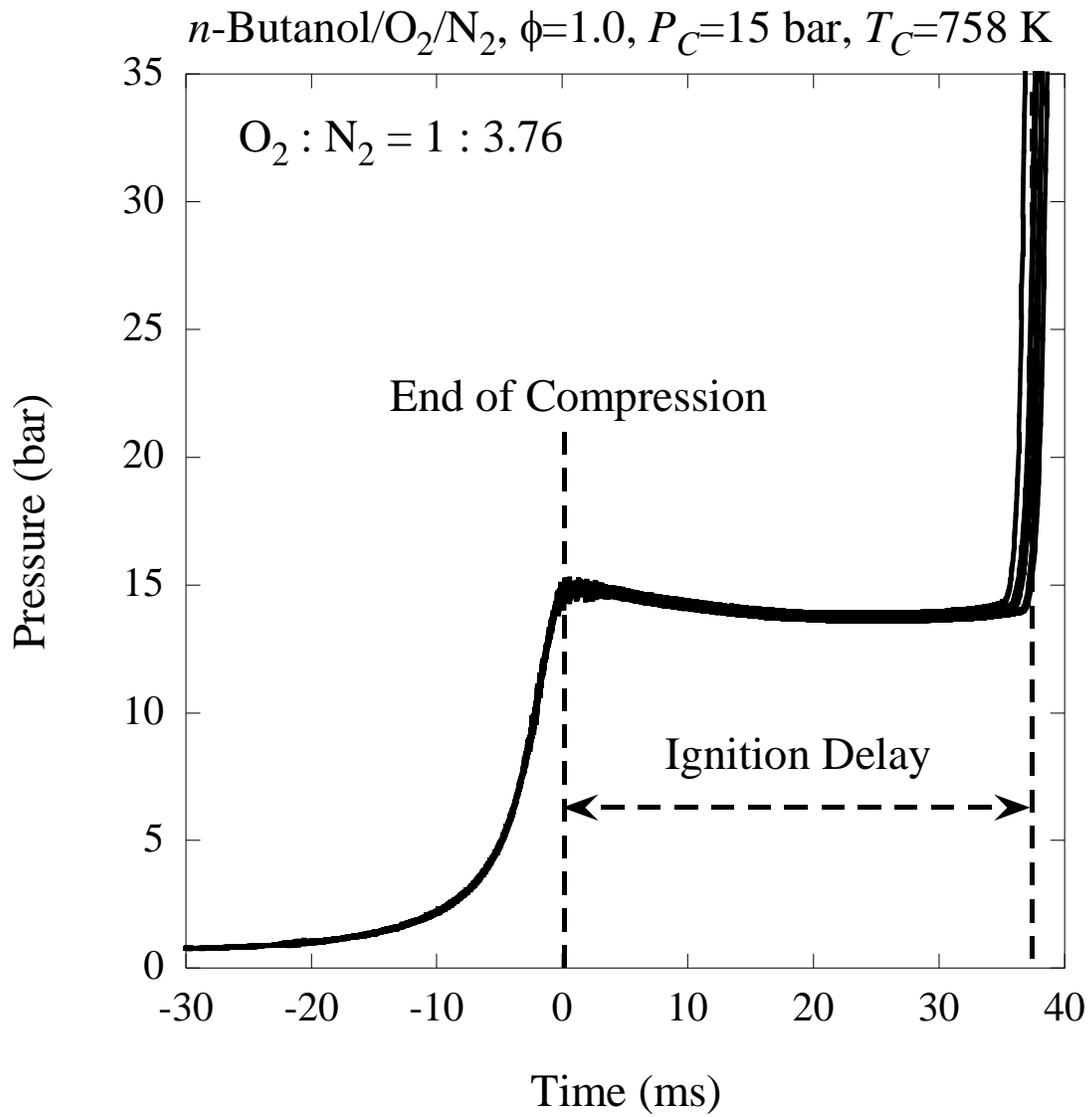

Figure 2

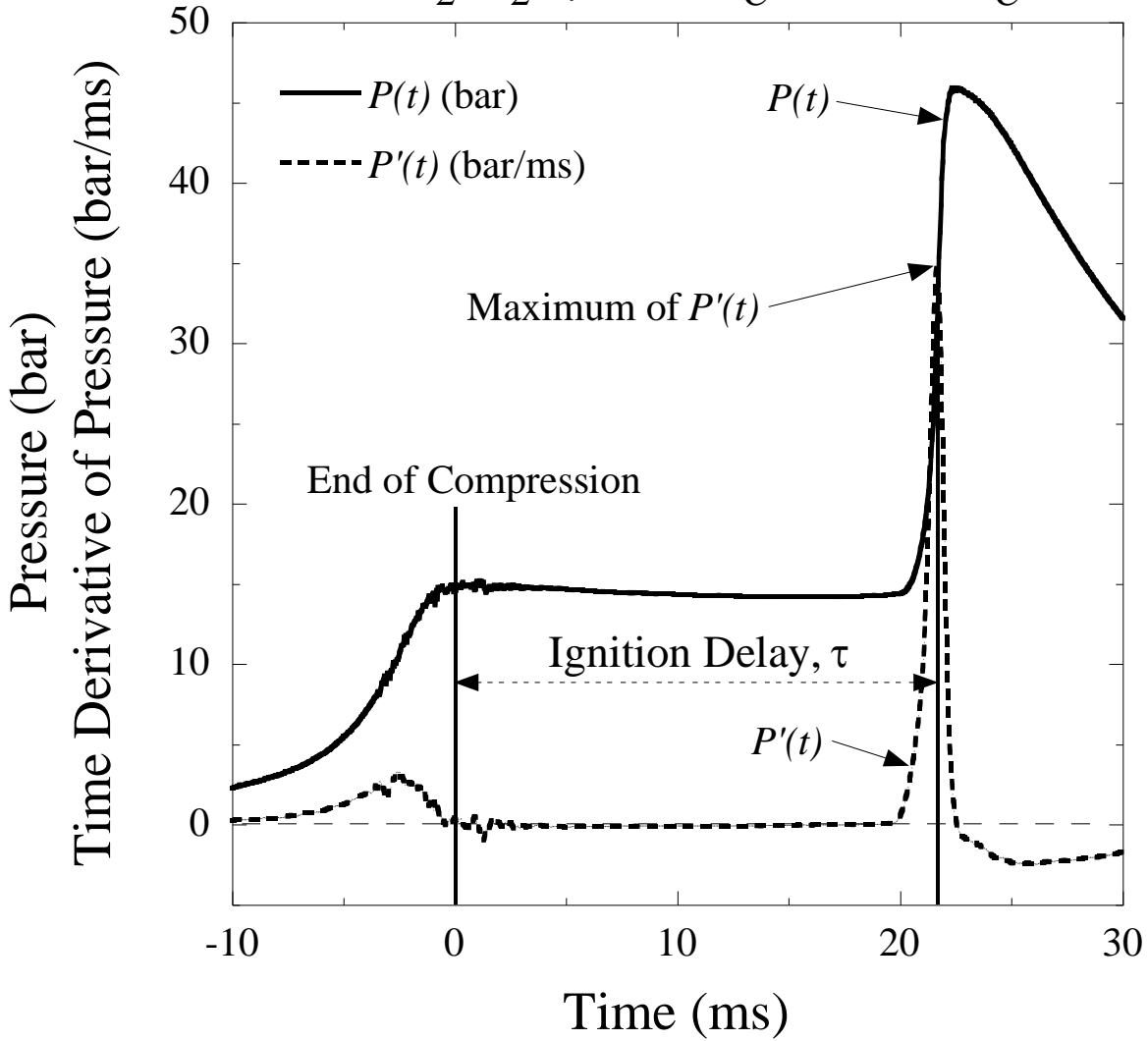

Figure 3

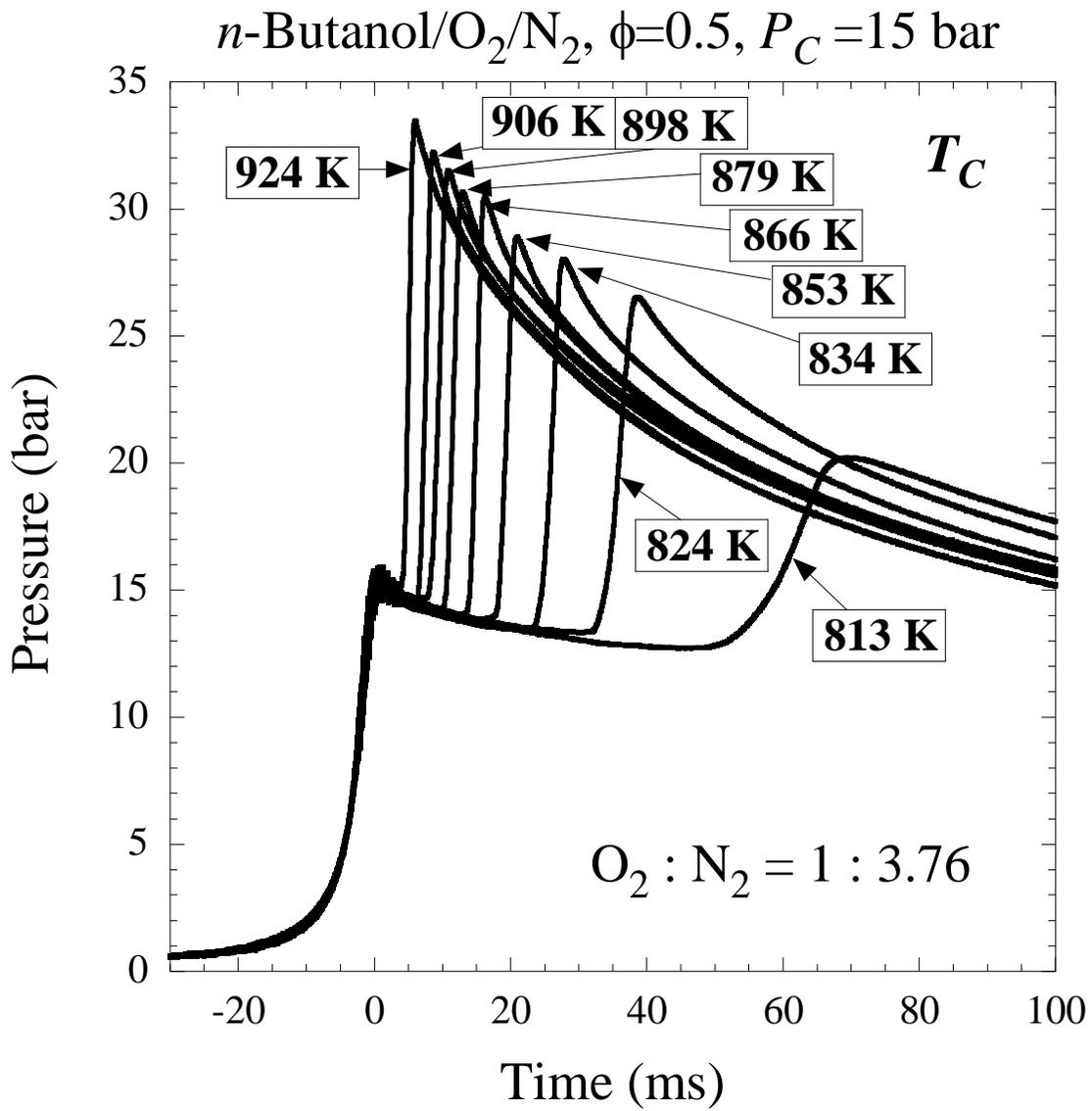

Figure 4

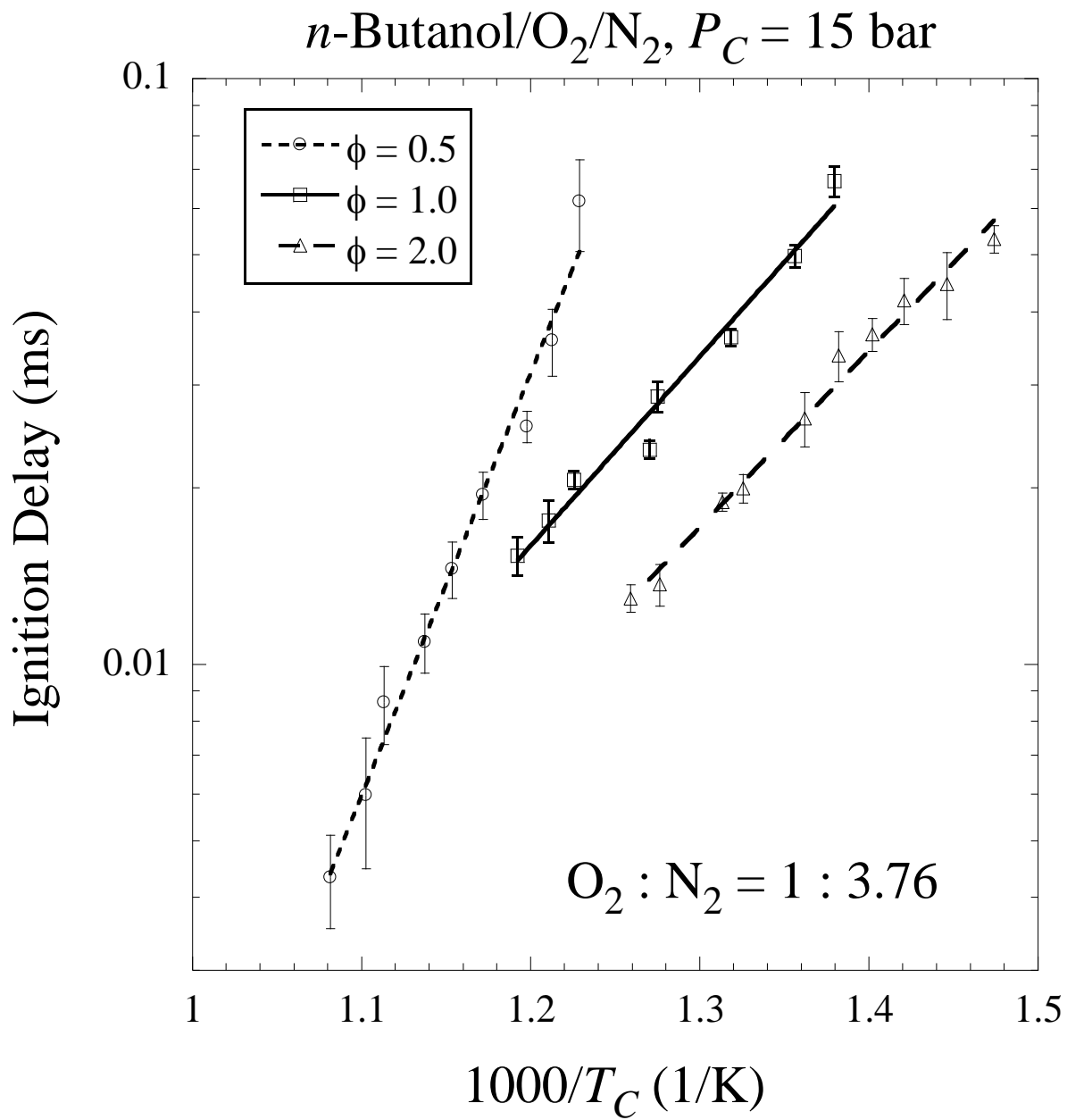

Figure 5

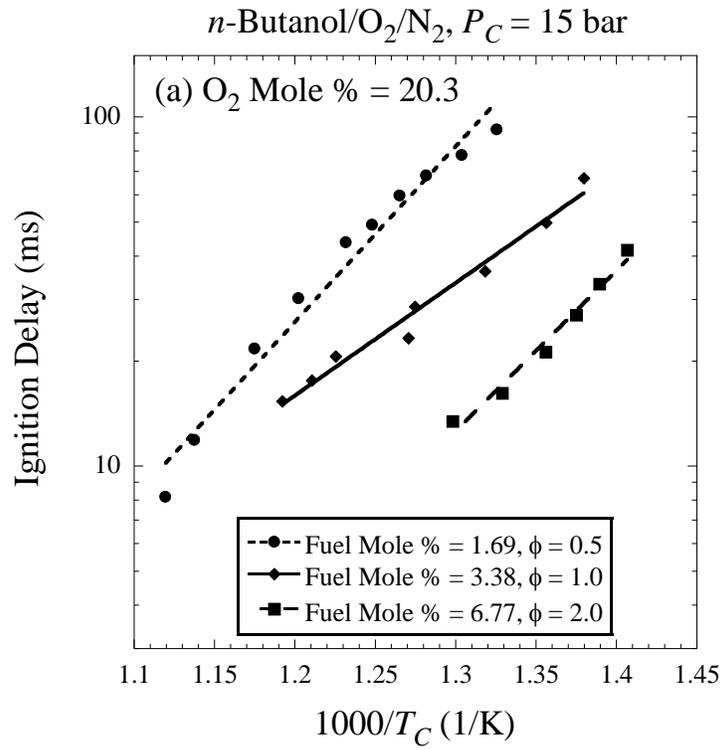

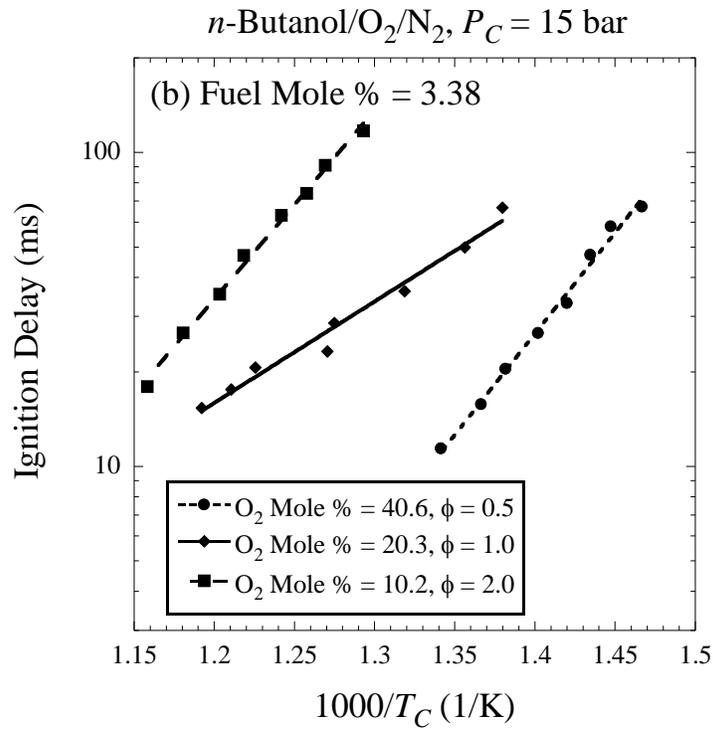

Figure 6

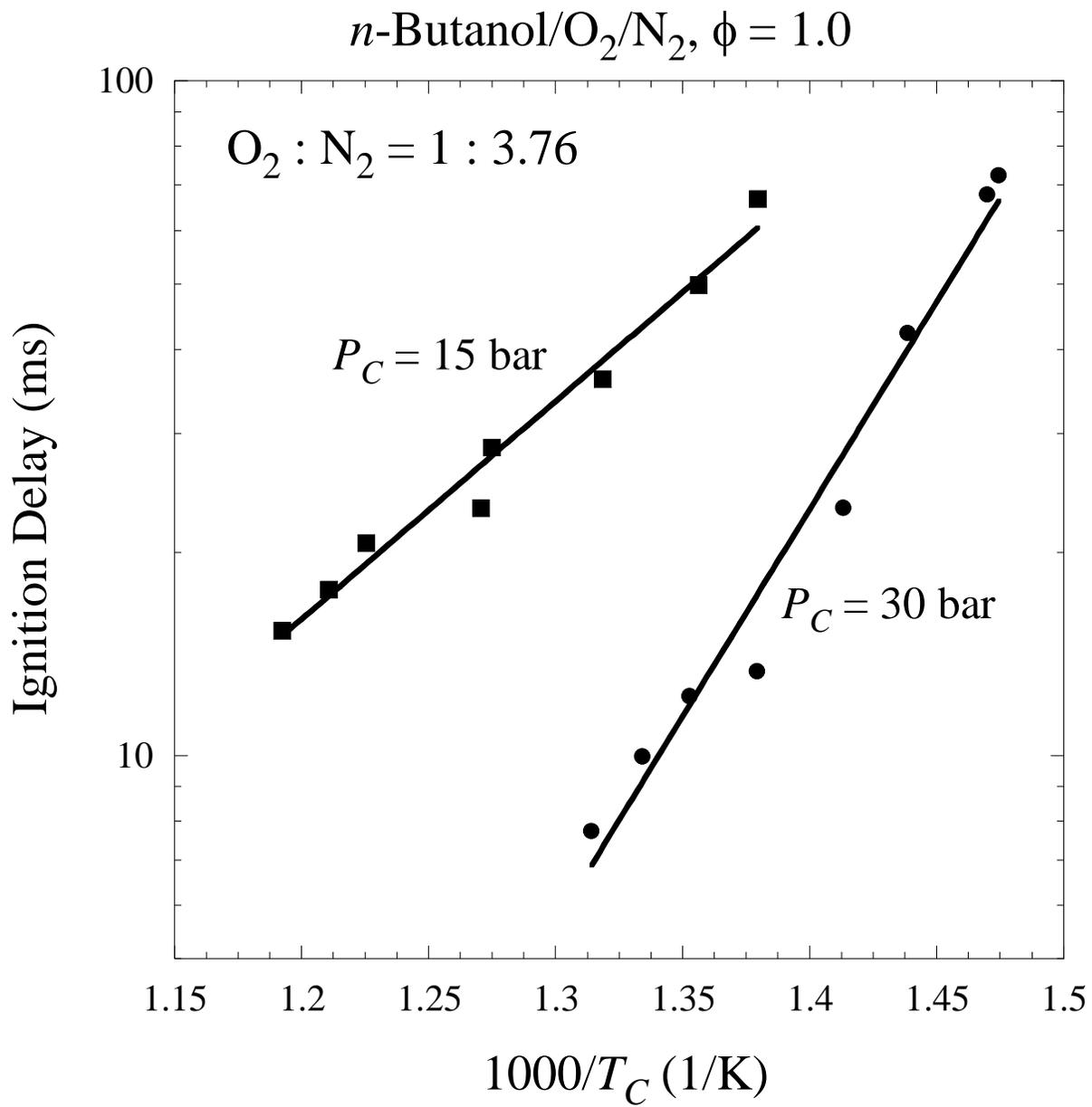

Figure 7

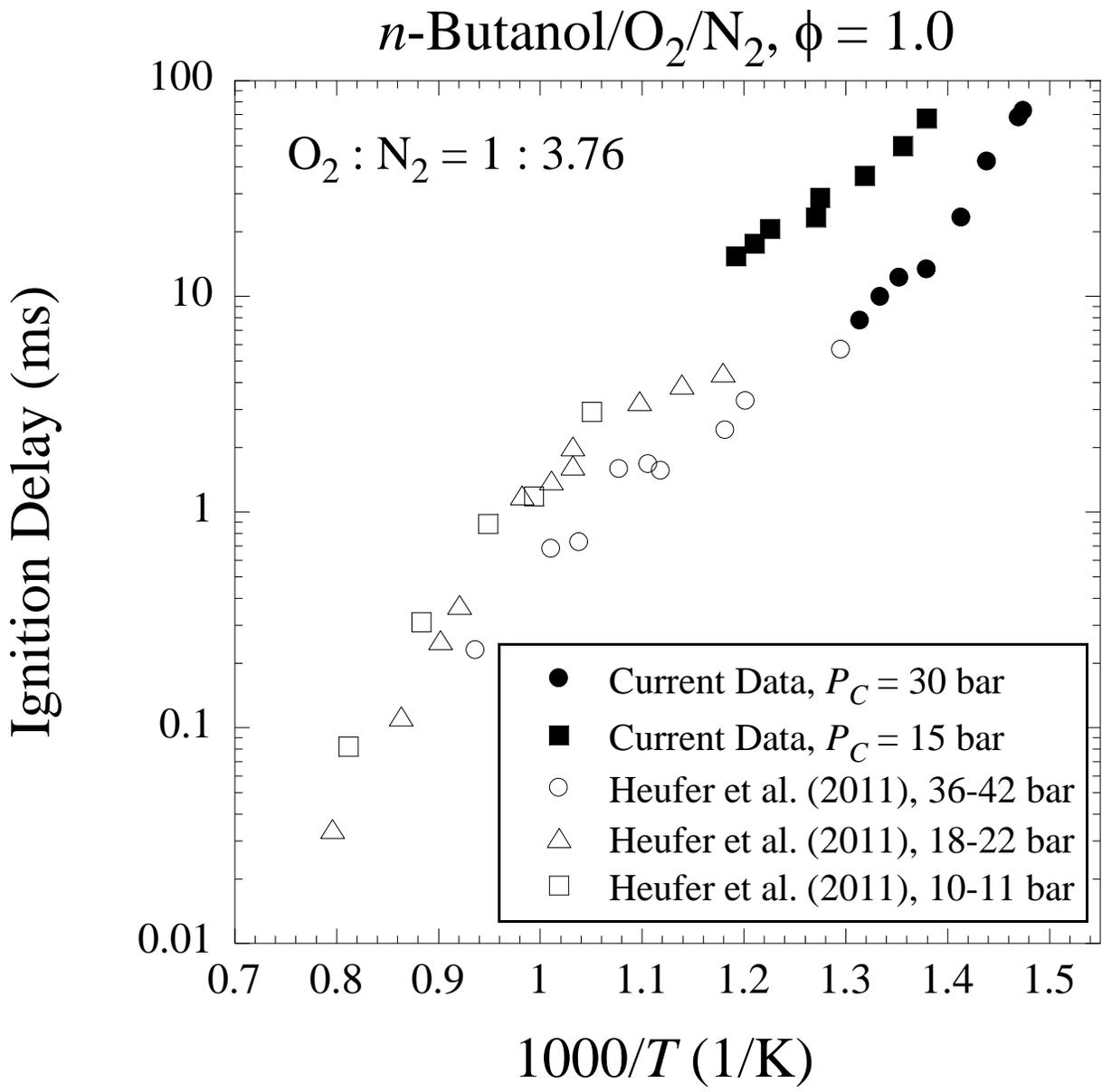

Figure 8

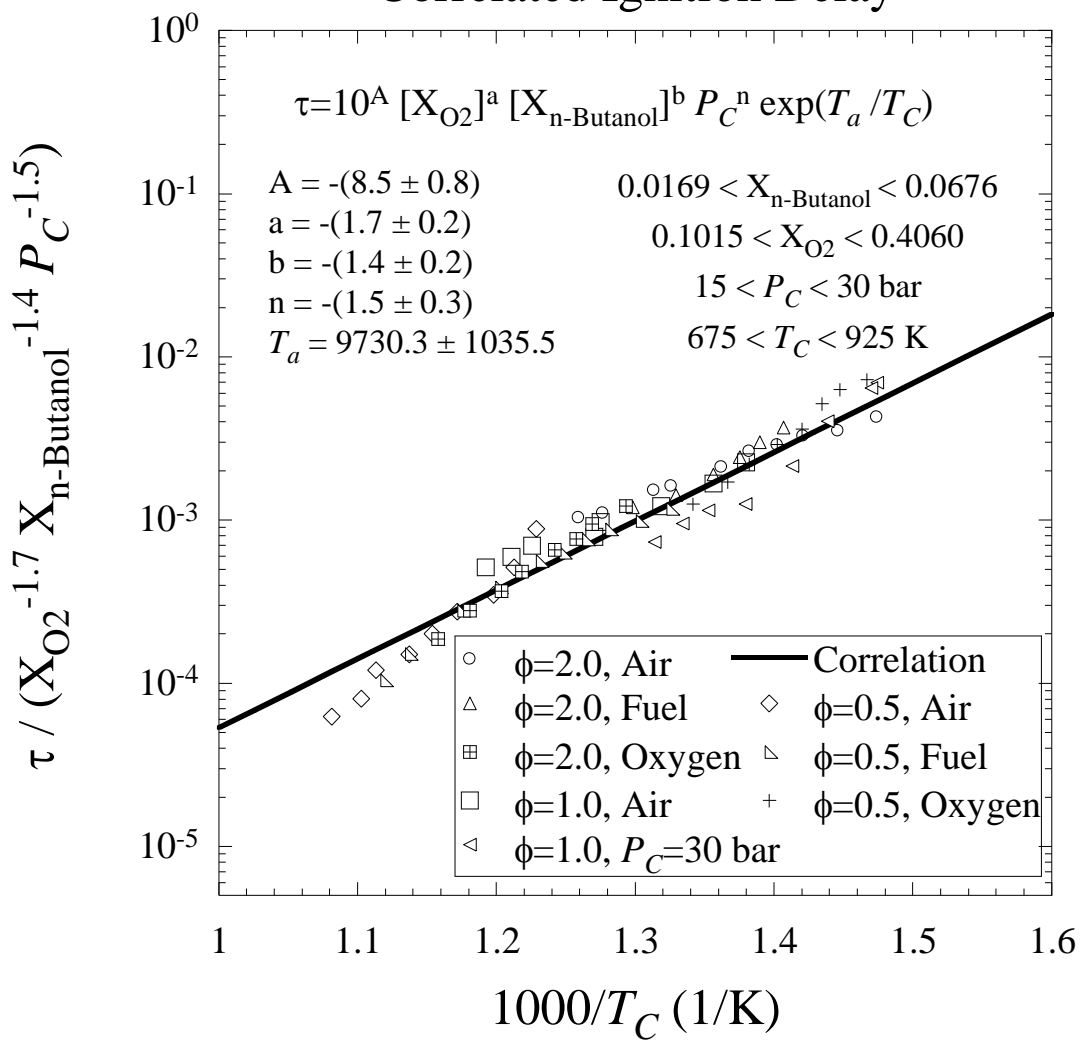

Figure 9

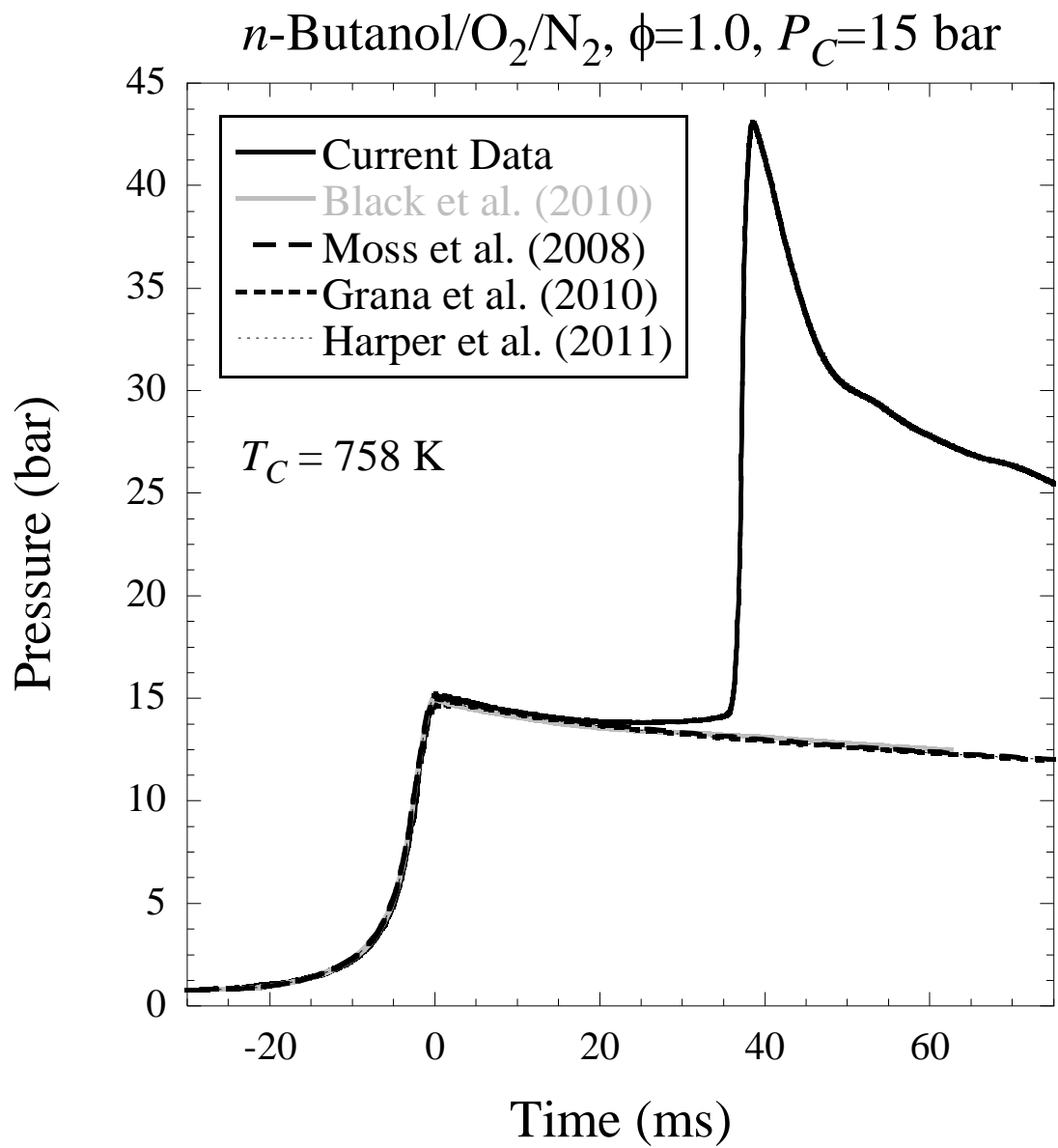

Figure 10

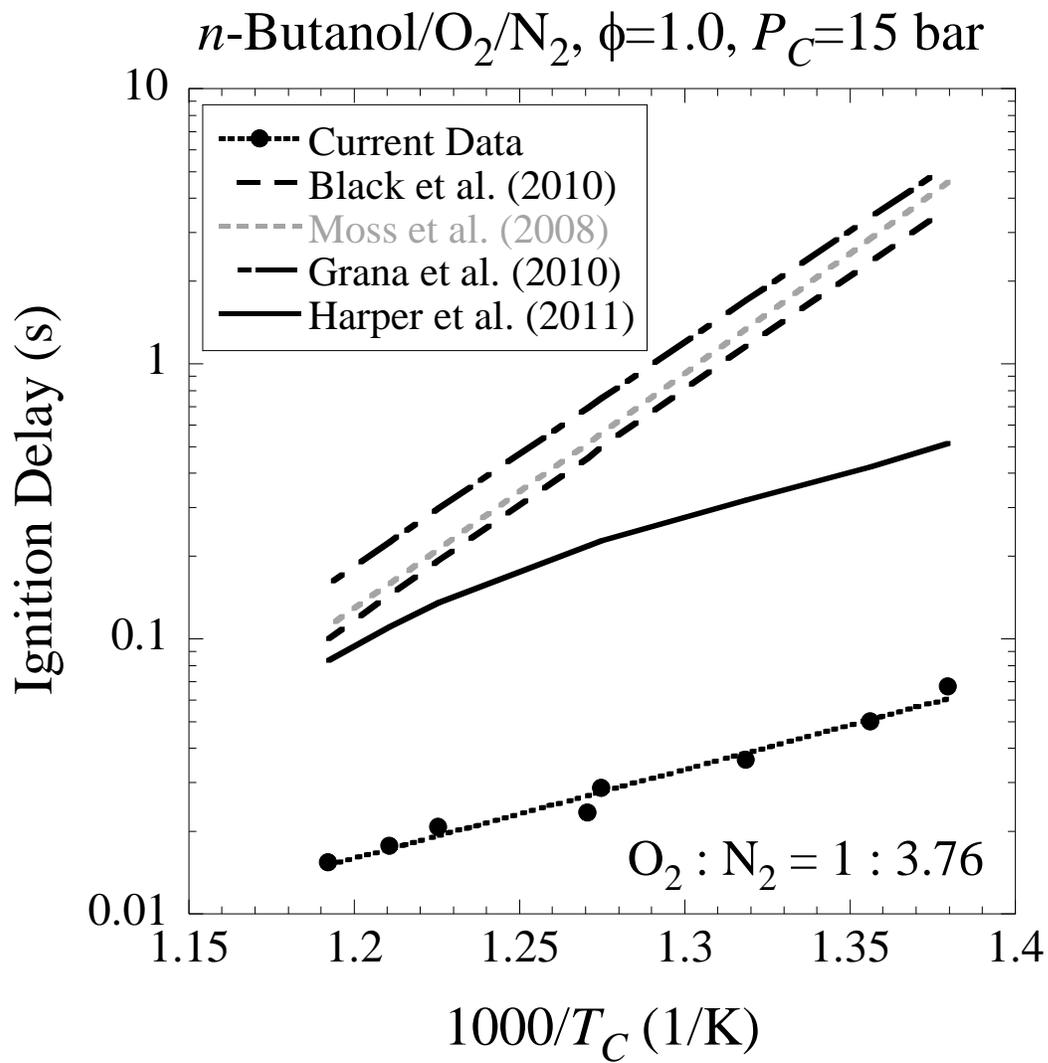

Figure 11

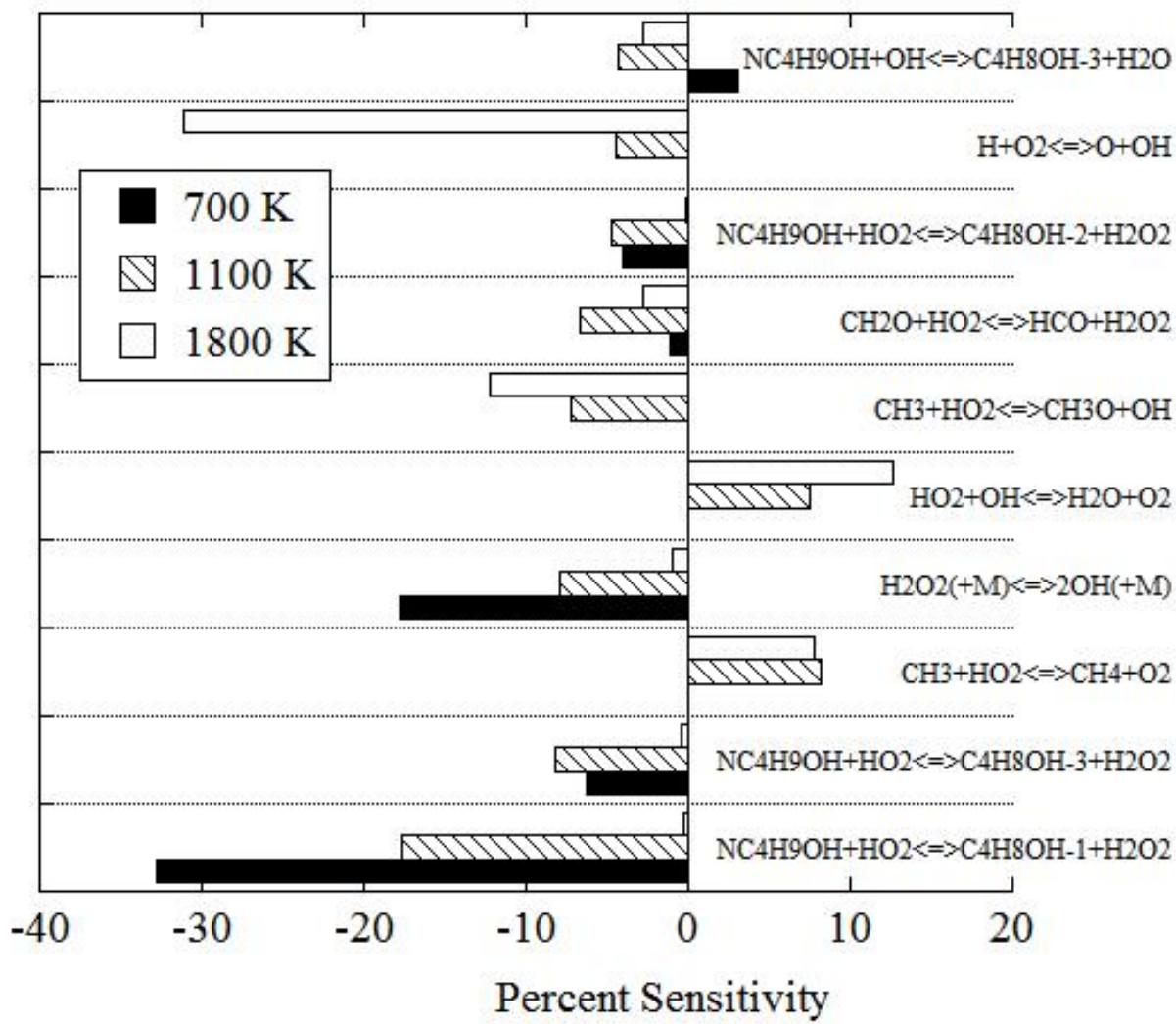

Figure 12

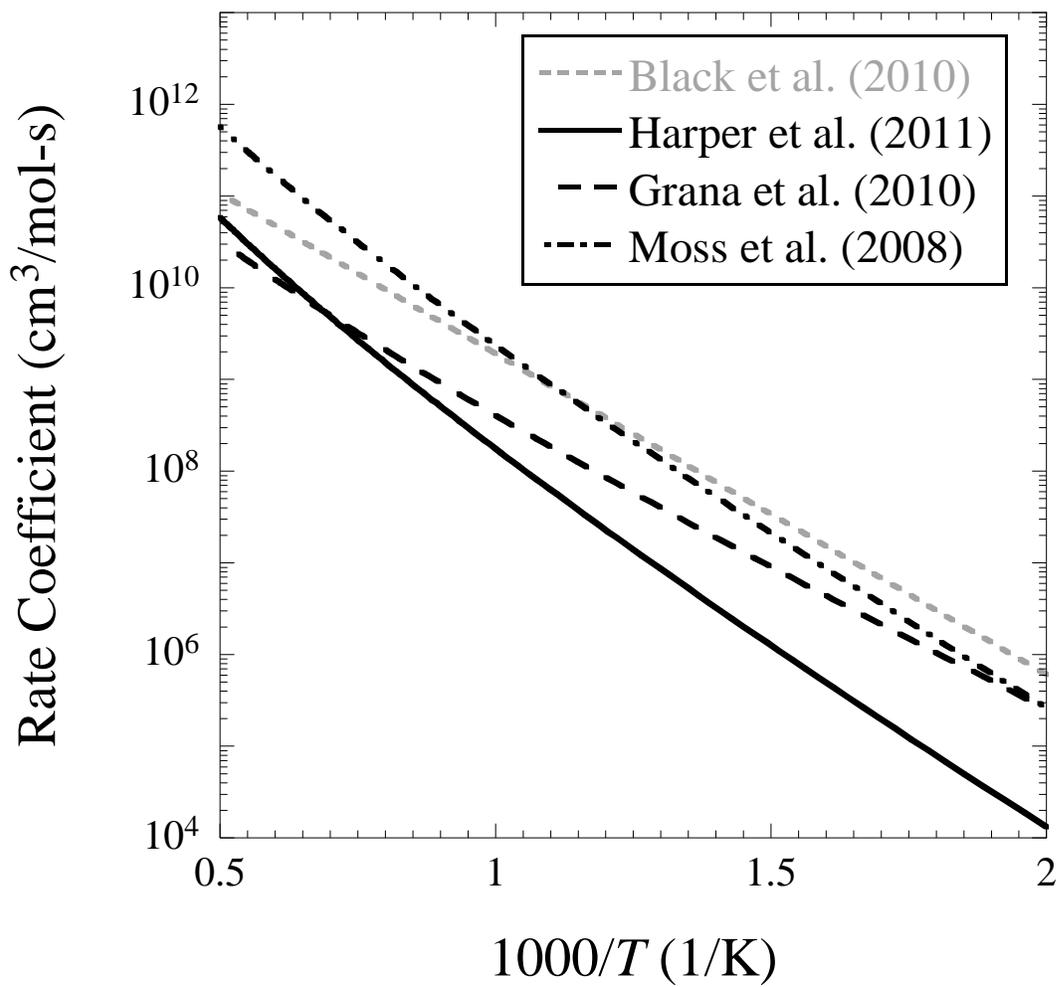

Figure 13

Figure 14

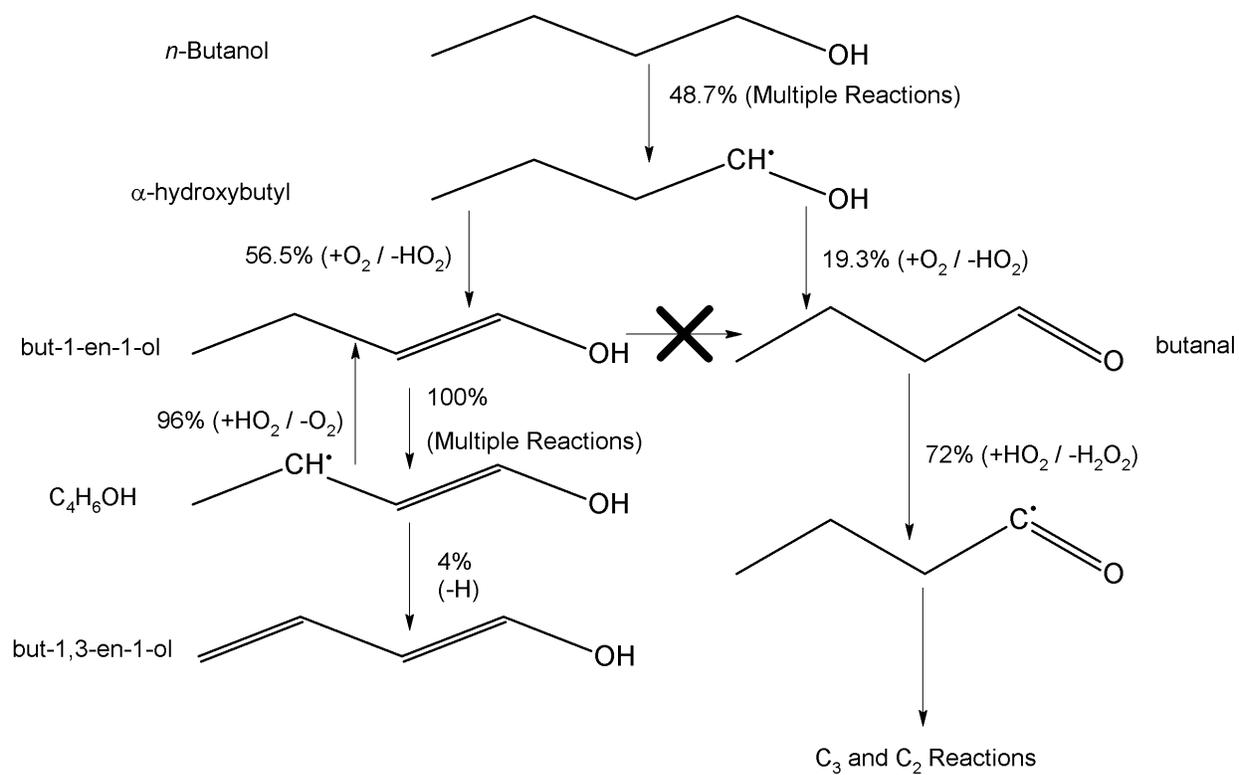

Figure 15

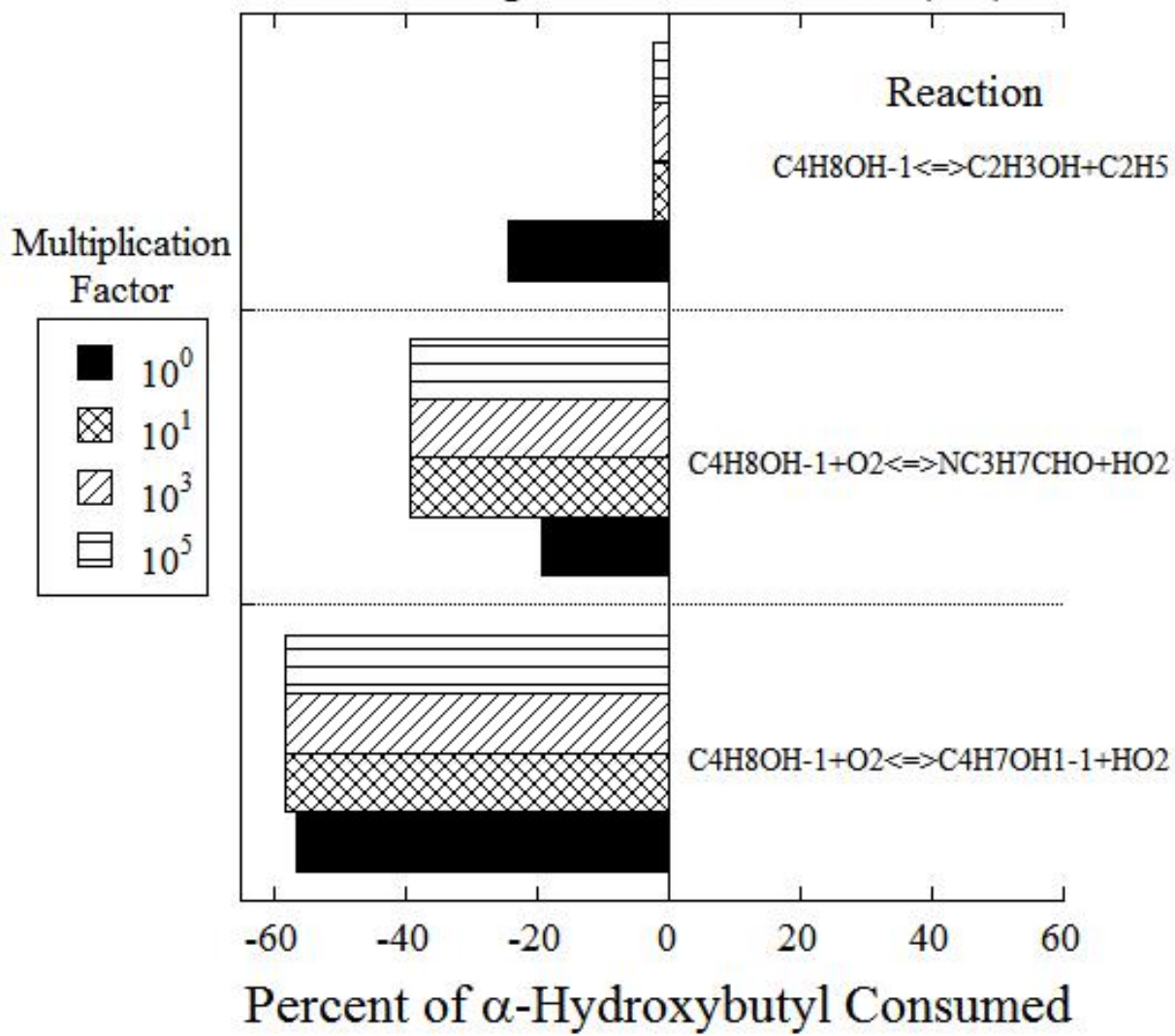

Figure 16

| Mole Percentage (%) | | | | |
|---|---|---|---|---|
| n-Butanol | $O_2$ | $N_2$ | $\phi$ | $P_C$ (bar) |
| 3.38 | 20.30 | 76.32 | 1.0 | 15 |
| 3.38 | 20.30 | 76.32 | 1.0 | 30 |
| 1.72 | 20.65 | 77.63 | 0.5 | 15 |
| 6.54 | 19.63 | 73.82 | 2.0 | 15 |
| 3.38 | 40.60 | 56.02 | 0.5 | 15 |
| 3.38 | 10.15 | 86.46 | 2.0 | 15 |
| 1.69 | 20.30 | 78.01 | 0.5 | 15 |
| 6.76 | 20.30 | 72.94 | 2.0 | 15 |

Table 1